\documentclass[]{interact}

\usepackage{epstopdf}% To incorporate .eps illustrations using PDFLaTeX, etc.
\usepackage[caption=false]{subfig}% Support for small, `sub' figures and tables
\usepackage{listings}%
\usepackage{float}
\usepackage[T1]{fontenc}
\usepackage{lmodern}
\usepackage{textcomp}
\usepackage{algorithm}
\usepackage{algorithmic}
\usepackage{placeins}

\usepackage[numbers,sort&compress]{natbib}% Citation support using natbib.sty
\bibpunct[, ]{[}{]}{,}{n}{,}{,}% Citation support using natbib.sty
\renewcommand\bibfont{\fontsize{10}{12}\selectfont}% Bibliography support using natbib.sty

\theoremstyle{plain}% Theorem-like structures provided by amsthm.sty

\theoremstyle{definition}

\theoremstyle{remark}

\begin{document}

%\articletype{Application Note}% Specify the article type or omit as appropriate

\title{Sparse-Group Boosting with Balanced Selection Frequencies: A Simulation-Based Approach and R Implementation}

\author{
\name{Fabian Obster \textsuperscript{a} \textsuperscript{b} \thanks{CONTACT Fabian Obster Email: fabian.obster@unibw.de} and Christian Heumann\textsuperscript{b}}
\affil{\textsuperscript{a}Department of Business Administration, University of the Bundeswehr Munich, Werner-Heisenberg-Weg 39, Bavaria, Germany \\ \textsuperscript{b}Department of Statistics, LMU Munich, Ludwigstr. 33, Bavaria, Germany}
}

\maketitle

\begin{abstract}
This paper introduces a novel framework for reducing variable selection bias by balancing selection frequencies of base-learners in boosting and introduces the \texttt{sgboost} package in R, which implements this framework combined with sparse-group boosting. The group bias reduction algorithm employs a simulation-based approach to iteratively adjust the degrees of freedom for both individual and group base-learners, ensuring balanced selection probabilities and mitigating the tendency to over-select more complex groups. The efficacy of the group balancing algorithm is demonstrated through simulations. Sparse-group boosting offers a flexible approach for both group and individual variable selection, reducing overfitting and enhancing model interpretability for modeling high-dimensional data with natural groupings in covariates. The package uses regularization techniques based on the degrees of freedom of individual and group base-learners. Through comparisons with existing methods and demonstration of its unique functionalities, this paper provides a practical guide on utilizing sparse-group boosting in R, accompanied by code examples to facilitate its application in various research domains. 
\end{abstract}


\section{Introduction}

Regularized regression is used to model high-dimensional data to reduce the risk of overfitting and to perform variable selection. In many cases, covariates have natural groupings, such as with gene data or categorical data often found in survey data. In such cases, one may want to select whole groups of variables or just individual parts. Sparse-group boosting is a powerful statistical method that extends classical boosting methods through the incorporation of structured sparsity. This structure is particularly useful in high-dimensional settings where predictors exhibit natural groupings.
`sgboost` implements the sparse-group boosting in R and other useful functions for sparse group model interpretation unique to boosting, and visualization for group and individual variable selection. The package is available on CRAN \cite{obster_sgboost_2024}. \\
Despite their utility and predictive performance, existing variable selection techniques often lack explicit mechanisms to balance sparsity within and between groups, making sgboost as an implementation of sparse-group boosting a valuable contribution to the field of variable selection. \\
To address a well-known limitation in boosting methods—namely, the tendency to favor larger or more flexible groups due to inherent selection bias - we introduce a group balancing algorithm. This algorithm employs a simulation-based approach to iteratively adjust the degrees of freedom assigned to each group, thereby equalizing the selection probabilities under the null hypothesis of no association. By using the shrinkage of the $RSS$ through the degrees of freedom in ridge regression, the algorithm ensures that groups of differing sizes are penalized appropriately. This adjustment mitigates over-selection of complex groups and enhances the overall interpretability and fairness of the model. The resulting group balancing can be combined with sparse-group boosting and is integrated into sgboost via the \verb|balance()| function, representing a significant methodological advancement in achieving balanced variable selection across heterogeneous groups.
\\
The increasing availability of high-dimensional datasets such as in economics, climate research, and bioinformatics, where variable selection plays a crucial role and group structures are relevant, motivates the implementation of sgboost. Many real-world datasets contain naturally predefined groups, such as geographical regions in climate modeling, questionnaire sections in survey-based research or gene data. Traditional boosting methods struggle with these structures, either selecting too many irrelevant variables or failing to capture group-level effects. By explicitly incorporating the group structure through two-level sparsity, sgboost addresses these challenges while also enhancing predictive accuracy and model interpretability. \\
Sparse-group boosting \cite{obster_sparse-group_2024} is an alternative method to sparse-group lasso \cite{simon_sparse-group_2013}, employing boosted Ridge regression. Although there are many methods of variable selection, most focus on group selection only, e.g. \cite{meier_group_2008}, \cite{huang_group_2009} and \cite{zhou_group_2010}, or individual variable selection e.g. \cite{berk_forward_1980}, \cite{zhang_variable_2016} and \cite{buhlmann_boosting_2007}. However, it should be noted, that in some cases of group variable selection with overlapping groups, one could also end up with sparse-group variable selection. There are not many R packages implementing sparse-group variable selection methods. There is 'SGL' \cite{simon_sgl_2019} implementing \cite{simon_sparse-group_2013}, 'sparsegl' \cite{liang_sparsegl_2023} with a faster implementation of the sparse-group lasso as well as "grpreg" \cite{breheny_grpreg_2024} implementing the group exponential least absolute shrinkage and selection operator (GEL) \cite{breheny_group_2015}, the Composite minimax concave penalty (cMCP) \cite{breheny_penalized_2009} and the group bridge \cite{huang_group_2009}. \\

The goal of this paper is to provide a practical guide including the code on how to use sparse-group boosting in R and get the most out of the method. The code is presented within this manuscript and can also be found also on GitHub together with the used dataset (https://github.com/FabianObster/sgboost-introduction). \\

While the mboost package \cite{hothorn_mboost_2023} already allows for structured regularization, sgboost is specifically designed for structural sparsity and simplifies the application of sparse-group boosting through a dedicated formula constructor, enhanced visualization, and interpretability tools. \\
To demonstrate the relevance of sgboost, we focus on two types of datasets: simulated data designed to reflect real-world sparsity structures and real-world data where sparse-group boosting provides clear advantages. In climate economics, the willingness of farmers to take adaptive measures against climate change is affected by multiple interdependent factors such as climatic/weather patterns, market conditions, and agronomic factors. These independent variables can be sorted into natural groups  (e.g., climatic variables, economic indicators, and agronomic environment), making such data ideal for sparse-group boosting. This paper explores how sgboost can be applied and performs in these settings, highlighting its practical benefits.
\section{Methods}
Throughout this paper, we consider a grouped dataset $X \in \mathbb{R}^{n \times p}$ with $G$ groups and each group $g$ contains $p_g$ variables. We also consider a generalized linear regression setting, such as explained in \cite{wood_generalized_2017} and \cite{mccullagh_generalized_1993} and their extended variations for ridge regression with a similar notation as in \cite{wieringen_lecture_2023}. For simplicity, we primarily illustrate the method using linear ridge regression; however, the approach can be readily extended to generalized linear models. We refer to the parameter vector as $\beta$ and to its estimate as $\widehat{\beta}$. The linear predictor with the response function $g^{-1}(\cdot)$ together form the conditional expectation of the response variable $\mathbb{E}[y|X] = \mu = g^{-1}(X\beta)$. As we can subset the design matrix, we can estimate a model using a subset of the design matrix, which we denote as $\widehat{\beta}_{V_g}$. Note that with this notation $\widehat{\beta}_{V_g}$ is not the same as the subset of $\widehat{\beta}$ using the index set ${V_g}$, but rather separate estimates. We will also use the notion of a Ridge hat matrix as defined in \cite{wieringen_lecture_2023} using the penalty term $\lambda \geq 0$: $H^\lambda = X(X^TX)^{-1}X^T$ and the degrees of freedom for which we use the definition of $\text{df}(\lambda) = \text{tr}(2H^\lambda- {(H^\lambda)}^2)$.
\subsection{Sparse-Group Boosting Framework}
sgboost extends traditional boosting by incorporating structured sparsity, combining component-wise and group-wise selection. This approach is particularly beneficial in high-dimensional settings such as correlated independent variables. Unlike standard boosting, which selects individual variables in isolation, sgboost allows the selection of groups or individual variables. This way, entire groups of predictors can be included or excluded, improving interpretability and predictive performance. The same holds for individual variables. \\
We define $p+G$ candidate sets denoted as $(V_l \subseteq \{ 1,..., p\})_{l \leq p+G}$. Each candidate set describes the indices of the variables to be considered as one group. This yields $p+G$ submatrices of the design matrix $X_{V_l}$ only containing the columns corresponding to the index set. 
\begin{itemize}
    \item The first $p$ are individual base-learners only containing one variable:
    \begin{equation*}
       V_l = \{l\} \text{ for } l \leq p,
    \end{equation*}
\item  and the remaining $G$ are group base-learners with group size $p_l$:
\begin{equation*}
     V_l = \{{(v_l)}_{1},..., {(v_l)}_{p_l} \} \subseteq \{1,...,p\}, \text{ for } l > p:
\end{equation*}
\end{itemize}
Through $V_l$, $l \geq p$, the group structure is defined using no overlapping groups.
Through this bi-level structure, within-group sparsity and between-group sparsity can be balanced. By using Ridge Regression, regularization is controlled through the degrees of freedom constraint, regulating sparsity levels.
The optimal base-learner is selected based on the residual sum of squares (RSS), yielding the most informative structure either via a group or individual variable at each step.
\begin{algorithm}
\caption{Sparse-Group $L^2$ Boosting Algorithm}
\begin{algorithmic}[1]
\STATE \textbf{Initialize:} $m \gets 0$, $\widehat{\beta}^{[0]} \gets \textbf{0}_p$, $\widehat{\mu}^{[0]} \gets X\widehat{\beta}^{[0]}$
\WHILE{$m < M$}
    \STATE $m \gets m+1$
    \FOR{each candidate set $V_l$, $l \leq p+G$}
        \STATE Compute residuals: $\widehat{u}^{[m-1]} \gets y - \widehat{\mu}^{[m-1]}$
        \STATE Fit Ridge regression:
        \begin{equation*}
            \widehat{\overline{\beta}}^{[m]}_{V_l} = ((X_{V_l})^T X_{V_l}+\lambda_l I_p)^{-1}(X_{V_l})^T(\widehat{u}^{[m-1]})
        \end{equation*}
        \STATE Set regularization parameter $\lambda_l$:
        \begin{equation}
        \lambda_l = 
        \begin{cases} 
            \lambda_l: \text{df}(\lambda_l) = \text{tr}(2H^\lambda_{V_l}-({H^\lambda_{V_l}})^2) = \alpha, & l \leq p \\
            \lambda_l: \text{df}(\lambda_l) = \text{tr}(2H^\lambda_{V_l}-({H^\lambda_{V_l}})^2) = 1-\alpha, & l > p
        \end{cases}
        \end{equation}
    \ENDFOR
    \STATE Select candidate set:
    \begin{equation*}
        l^* = \arg\min_{l \leq L} (\widehat{u}^{[m-1]} - X_{V_l} \widehat{\overline{\beta}}_{V_l})^T (\widehat{u}^{[m-1]} - X_{V_l} \widehat{\overline{\beta}}_{V_l})
    \end{equation*}
    \FOR{all $l \leq p+G$}
        \STATE Update coefficients:
        \begin{equation*}
        \widehat{\beta}^{[m]}_{{V_l}} =
        \begin{cases} 
            \widehat{\beta}^{[m-1]}+\nu \widehat{\overline{\beta}}_{V_{l^*}}, & l = l^* \\
            \widehat{\beta}^{[m-1]}, & l \neq l^*
        \end{cases}
        \end{equation*}
    \ENDFOR
    \STATE Update estimate:
    \begin{equation*}
        \widehat{\mu}^{[m]} = X \widehat{\beta}^{[m]}
    \end{equation*}
\ENDWHILE
\STATE \textbf{Output:} $\widehat{\beta}^{[M]}$
\end{algorithmic}
\end{algorithm}

The same algorithm can be used to fit (generalized) linear models by replacing the $L^2$ loss function with the modified loss function $\mathcal{L}$. This yields for individual base-learners $l \leq p$
\begin{align*}
   \mathcal{L} ^{[m]}_{V_l} = -\sum_{i=1}^n \ell_i(\widehat{\beta}^{[m-1]} + \beta_{V_l}) + \alpha \lambda_l  (\beta_{V_l})^T\beta_{V_l},
\end{align*}
and for group base-learners $l > p$
\begin{align*}
   \mathcal{L}^{[m]}_{V_l} = -\sum_{i=1}^n \ell_i(\widehat{\beta}^{[m-1]} + \beta_{V_l}) + (1-\alpha)\lambda_l  (\beta_{V_l})^T\beta_{V_l}.
\end{align*}
\FloatBarrier
\subsection{Group adjustment}
Variable selection bias can occur in the presence of grouped variables, such as categorical or functional data, making the definition of the degrees of freedom $df(\lambda) = \text{tr}(2H^\lambda- (H^\lambda)^2)$ preferable \cite{hofner_framework_2011}. However, group selection bias can still occur because of the group size. The same issue occurs in the sparse group lasso of the group bridge, which is met by using group standardization depending on the type of regularization, such as $\sqrt{p_g}$ \cite{simon_sparse-group_2013} \cite{breheny_penalized_2009}. Many algorithms use such an adjustment, which is also referred to as outer adjustment \cite{buch_systematic_2023}. This is to prevent an over-selection of groups with larger group sizes. Unlike traditional methods that use fixed penalties or shrinkage parameters, this algorithm dynamically adjusts selection probabilities through repeated sampling, enabling data-driven balancing. To overcome this issue, we introduce a simulation-based algorithm that balances the selection chance of one group over another by using the degrees of freedom. 

Assume we have $G$ groups, described by the index sets $V_1, ..., V_G$, with group sizes $p_1, ..., p_g$. Denote the scaling vector for the degrees of freedom as $d = (d_1,...,d_G)$, where each group $g$ has its own value for the degrees of freedom $d_g$ for $g \leq G$.  
\begin{algorithm}
\caption{Group balancing algorithm}
\begin{algorithmic}[2]
\STATE \textbf{Initialize:} Set $r = 0$ and $d_g^*= d_g^{[1]} \equiv c$ with constant $c \in ]0,1[$ for all $g \leq G$. A reasonable starting value is $c=0.5$.
\FOR{$r \leq R$}
    \STATE $r \gets r+1$
    \STATE Simulate $K$ versions of the outcome variable $y^{(k)}$, e.g., $y^{(k)} \sim \mathcal{N}(0_n,I_n)$ for $k \leq K$.
    \FOR{$k \leq K$}
        \STATE Fit the learning algorithm $f: X \to y^{(k)}$ with the degrees of freedom $d^*$ to obtain the fitted model $\widehat{f}^{(k)}$.
    \ENDFOR
    \STATE Retrieve the activation vector $\big(s^{(k)}_1, ..., s^{(k)}_G\big) \in \{0,1\}^G$ for each $\widehat{f}^{(k)}$, indicating selected groups. If a group is selected, the value one is assigned; if not, then the value zero.
    \STATE Compute the average selection frequency vector:
    \[
    \overline{s} = \Big(\overline{s}_1,...,\overline{s}_G\Big) = \Big(\frac{1}{K}\sum_{k = 1}^K s^{(k)}_1, ..., \frac{1}{K}\sum_{k = 1}^K s^{(k)}_G\Big)^T.
    \]
    \STATE Compute the error vector:
    \[
    c^{[r]} = \big(\frac{1}{G},..., \frac{1}{G}\big)^T-\overline{s}.
    \]
        \FOR{$g \leq G$}
            \STATE Update:
            \IF{$\sum_{g=1}^G ( c_g^{[r]})^2 < \sum_{g=1}^G ( c^*_g)^2$}
            \STATE 
            \begin{align*}
            d_g^* &= d_g^{[r-1]} \\
            d_g^{[r]} &=  d_g^* + \nu c^{[r]}
            \end{align*}
            \ELSE 
            \STATE 
            \begin{align*}
            \nu &= \gamma \nu \\
            d_g^{[r]} &= 
                (1-\eta) d_g^* + \eta(d_g^{[r-1]} + \nu c^{[r]}). 
            \end{align*}
            \ENDIF
        \ENDFOR
\ENDFOR
\STATE \textbf{Return:} $d^*$ as the degrees of freedom scaling vector.
\end{algorithmic}
\end{algorithm}
\FloatBarrier
Note that it is sufficient to run the boosting algorithm for only one step instead of fitting the whole algorithm, to achieve the balance, as the algorithm does not depend on the actual outcome variable. This allows for a highly efficient estimation of group preference without fully fitting the model. This compensates for simulating many repetitions of the outcome variable and refitting for each sample  \\ The general idea is to decrease the degrees of freedom for over-selected groups and increase the degrees of freedom for under-selected groups. The step size is proportional to the imbalance, meaning strongly imbalanced groups are adjusted more than slightly imbalanced groups, and is multiplied by the learning rate $\nu$, which impacts how far the update goes away from the current estimate $d^*$. A larger $\nu$ leads to larger corrections, hence fewer necessary iterations, but may cause oscillations or overshooting, especially in small-sample or highly collinear settings. Choosing an appropriate learning rate is therefore a trade-off between speed of convergence and stability. Also, $K$, the number of samples of the outcome variable, increases stability and should also affect the choice of the learning rate. If the algorithm overshoots and the overall imbalance increases, a convex combination between the current best estimate and the updated parameter is used, where $\eta$ is the mixing parameter. Also, the learning rate is reduced by $ \gamma \in ]0,1[$, e.g. 0.9 to avoid overshooting in future steps. This approach incorporates new information while preserving the information from the previous best estimate, balancing exploration and robustness. The algorithm can be stopped after a fixed number of iterations $R$ or if $\sum_{g=1}^G ( c_g^{[r]})^2$ is smaller than some predefined value. The algorithm does not yield a unique solution, which depends strongly on the initialization of $c$. The existence of a solution that balances the selection frequencies is guaranteed by the mean-value theorem and the law of large numbers if $K \to \infty$. One can verify that $\text{df}(\lambda_g) \to 0$ implies $\overline{s}_g \to 0$ and $\text{df}(\lambda_g) \to 1$ implies $\overline{s}_g \to z>0$. Furthermore, the residual sum of squares is monotonous and continuous in $df(\lambda)$. Therefore, $\frac{1}{K}\sum_{k = 1}^K s^{(k)}_g = P(RSS(\lambda_g) = max_{l \leq G}(RSS(\lambda_l))$ is also monotonous and continuous in $\text{df}(\lambda_g)$. A unique solution could be achieved by fixing the degrees of freedom for one base-learner and only updating the others. Another variation of the algorithm is to perform the algorithm by updating the ridge regularization parameter $\lambda$ for each group instead of the degrees of freedom. The algorithm can be used for group boosting, but also for sparse-group boosting by expanding the group index set to include also $ \tilde{V} = {1},..., {p}, V_1,...V_G$, leading to overlapping groups. In this case, one could also update the error vector to $c^{[r]} = \big( \frac{\alpha}{p+G},\overset{^p}{\quad...\quad }, \frac{\alpha}{p+G}, \frac{1-\alpha}{p+G},\overset{^G}{\quad...\quad }, \frac{1-\alpha}{p+G}\big)^T-\overline{s}$, instead of mixing the degrees of freedom. In this case, $\alpha$ would have a natural interpretation, though the odds of an individual base-learner being selected over a group base-learner as $\frac{\alpha}{1-\alpha}$. This would make the choice of $\alpha$ much easier. Then, $\alpha = 0$ would still correspond to group boosting, $\alpha = 1$ to componentwise boosting. The case of $\alpha = 0.5$, would lead to equal selection frequencies of each base-learner, regardless of the group size and type of base-learner.
\section{Results}\label{sec2}
We first simulate the sample data and corresponding group structure with 40 equal-sized groups to show the sparse-group boosting workflow. Based on a linear regression model we simulate the response variable y as part of the data.frame with $n=100$ observations and $p=200$ predictor variables (each group is formed by 5 predictors).
\begin{lstlisting}
beta <- c(
  rep(5, 5), c(5, -5, 2, 0, 0), rep(-5, 5),
  c(2, -3, 8, 0, 0), rep(0, (200 - 20))
)
X <- matrix(data = rnorm(20000, mean = 0, sd = 1), 100, 200)
df <- data.frame(X) %>%
  mutate(y = X %*% beta+rnorm(100, mean = 0, sd = 1)) %>%
  mutate_all(function(x){as.numeric(scale(x))})
group_df <- data.frame(
  group_name = rep(1:40, each = 5), 
  variable_name = colnames(df)[1:200]
)
\end{lstlisting}
\subsection{Defining the model}
Now we use the group structure to describe the sparse group boosting formula with the function \verb|create_formula()|. We only need the data.frame() describing the group structure. It should contain two variables, one indicating the name of the variable in the modeling data (\verb|var_name|), and one indicating the group it belongs to (\verb|group_name|). Additionally, we need to pass the mixing parameter \verb|alpha| and the name of the outcome variable.
\begin{lstlisting}
sgb_formula <- create_formula(
  alpha = 0.4, group_df = group_df, outcome_name = "y",
  group_name = "group_name", var_name = "variable_name")
\end{lstlisting}
This function returns an R-formula consisting of $p$ model terms defining the individual base-learners and $G$ group base-learners. 
\begin{lstlisting}
labels(terms(sgb_formula))[[1]]
## bols(X1, df = 0.4, intercept = FALSE)
labels(terms(sgb_formula))[[201]]
## bols(X1, X2, X3, X4, X5, df = 0.6, intercept = FALSE)
\end{lstlisting}
\subsection{Fitting and tuning the model}
\verb|sgboost| is to be used in conjunction with the \verb|mboost| package, which provides many useful functions and methods that can also be used for sparse-group boosting models. \\
Now we pass the formula to \verb|mboost()| and use the arguments as seems appropriate.
The main hyperparameters are \verb|nu| and \verb|mstop|. For model tuning, the function
\verb|cvrisk| can be used and plotted. Running the cross-validation/bootstrap in parallel can speed up the process.
\begin{lstlisting}
sgb_model <- mboost(
  formula = sgb_formula, data = df,
  control = boost_control(nu = 1, mstop = 600)
)
cv_sgb_model <- cvrisk(sgb_model)
mstop(cv_sgb_model)
## 204
plot(cv_sgb_model
\end{lstlisting}

\begin{figure}[H]
\centering
\includegraphics[scale=0.5]{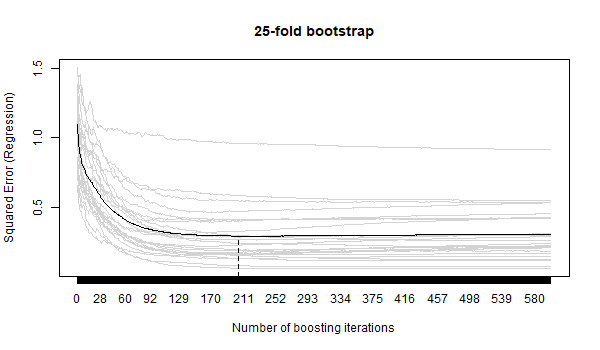}
\caption{Out of sample error depending on the boosting iteration}\label{fig:cv}
\end{figure}
In this example, the lowest out-of-sample risk is obtained at 204 boosting iterations, so we only use the first 204 updates for the final model. 
\subsection{Interpreting and plotting a sparse-group boosting model}
\verb|sgboost| has useful functions to understand sparse-group boosting models, and reflects that the final model estimates of a specific variable in the dataset can be
attributed to group base-learners as well as individual base-learners depending on the boosting iteration.

\subsubsection{Variable importance}
A good starting point for understanding a sparse-group boosting model is the 
variable importance. In the context of boosting, the variable importance can be 
defined as the relative contribution of each predictor to the overall reduction
of the loss function (negative log-likelihood).
\verb|get_varimp()| returns the variable importance of each base-learner/predictor 
selected throughout the boosting process. In the case of the selection of an individual variable - call it $x_1$ - as well as the group it belongs to -$x_1, x_2, ... x_p$ -, both base-learners (predictors) will have an associated
variable importance as defined before. This allows us to differentiate between the individual contribution of $x_1$ as its own variable and the contribution of the 
group $x_1$ belongs to. It is impossible to compute the aggregated variable importance of $x_1$ as it is unclear how much $x_1$ contributes to the group. However, the aggregated coefficients can be computed using \verb|get_coef()|, which also returns the aggregated importance of all groups vs. all individual variables in a separate data.frame. With \verb|plot_varimp()|
one can visualize the variable importance as a barplot. Since group sizes can be large, the function allows for cutting of the name of a predictor after \verb|max_char_length| characters.  One can indicate the maximum number of predictors to be printed through \verb|n_predictors| or through the minimal 
variable importance a predictor has to have through \verb|prop|. Through both parameters, the number of printed entries can be reduced. Note, that in this case, the relative importance of groups in the legend is based only on the
plotted variables and not the ones removed. Adding information about the direction of effect sizes, one could add arrows behind the bars \cite{obster_financial_2024}. For groups, one can use the aggregated coefficients from \verb|get_coef()|.

\begin{lstlisting}
slice(get_varimp(sgb_model =sgb_model_linear)$varimp,1:5)
\end{lstlisting}
\begin{verbatim}
# A tibble: 5  6
  reduction blearner        predictor   selfreq type  relative_
                                                      importance
      <dbl> <chr>           <chr>       <dbl>   <chr>     <dbl> 
1    0.297  bols(X1, X2,... X1, X2, ... 0.206   group     0.301 
2    0.288  bols(X18, in... X18         0.0196  indi...   0.292 
3    0.230  bols(X11, X1... X11, X12... 0.25    group     0.233 
4    0.0414 bols(X7, int... X7          0.0784  indi...   0.0419
5    0.0392 bols(X6, int... X6          0.0833  indi...   0.0397
\end{verbatim}
\begin{lstlisting}
get_varimp(sgb_model = sgb_model_linear)$group_importance
\end{lstlisting}
\begin{verbatim}
    # A tibble: 2 × 2
  type       importance
  <chr>           <dbl>
1 group           0.534
2 individual      0.466
\end{verbatim}
\begin{lstlisting}
plot_varimp(sgb_model = sgb_model_linear, n_predictors = 15)
\end{lstlisting}
\begin{figure}[H]
\centering
\includegraphics[scale=0.5]{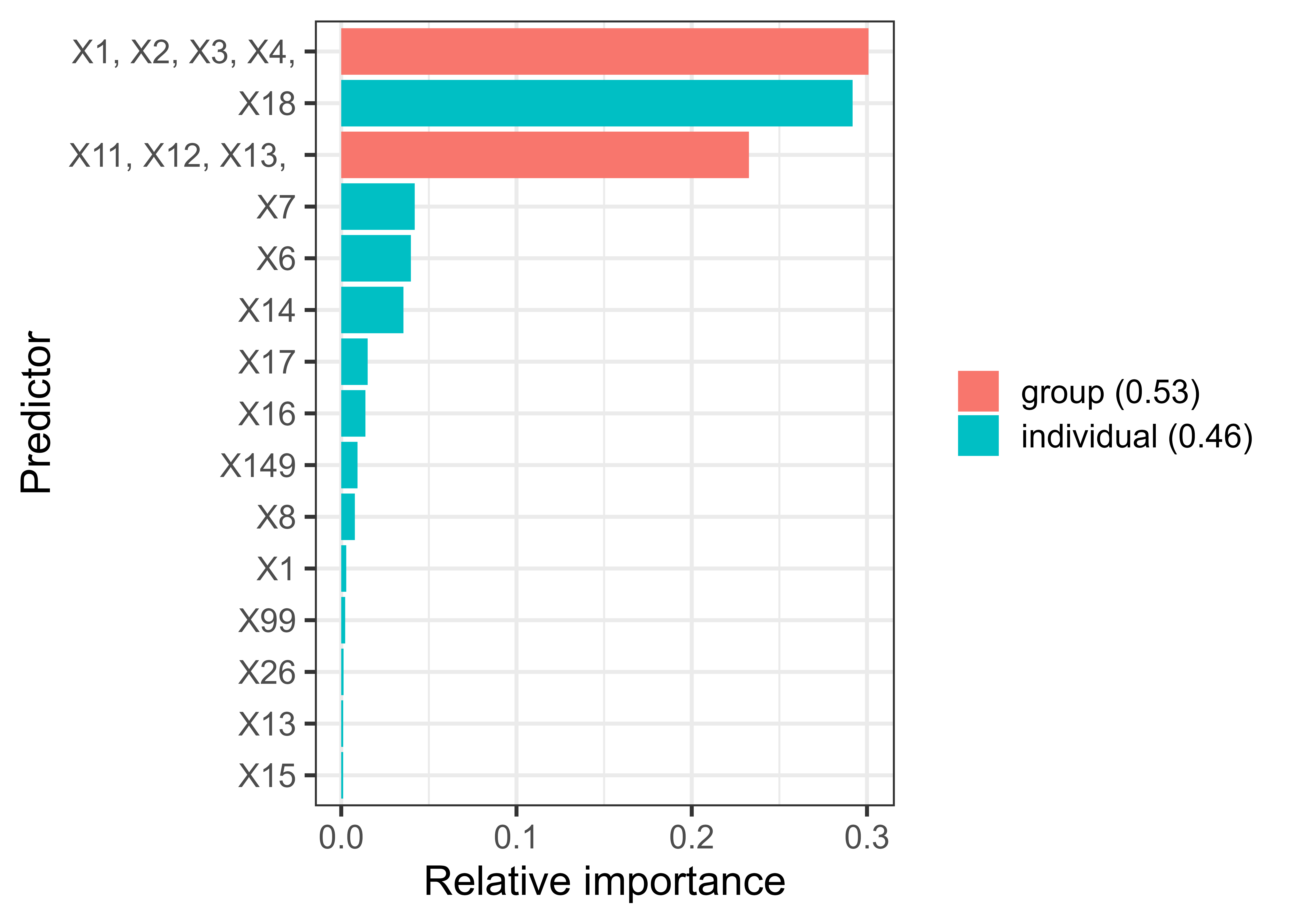}
\caption{Variable importance of the sparse-group boosting model for simulated data. The variable labels in the groups are cut off after 15 characters by defalut.}\label{fig:varimp}
\end{figure}
In this example, we see that both individual variables and groups were selected and contributed to the reduction of the loss function. The most important predictor is the first group, followed by variable 18, and then by group three. This is in line with what was simulated, as variable 18 has the biggest beta value, and groups one and three are full groups, meaning all variables within the groups have a non-zero beta coefficient. Groups two and four have within-group sparsity, therefore, they were selected as individual variables rather than groups.
\subsubsection{Model coefficients}
The resulting coefficients can be retrieved through \verb|get_coef()|
In sparse-group boosting models, a variable in a dataset can be selected as an individual variable or through a group. Therefore, there can be two associated effect sizes for the same variable. This function aggregates both and returns them in a data.frame
sorted by the effect size \verb|'effect'|.
\begin{lstlisting}
slice(get_coef(sgb_model = sgb_model)$raw,1:5)
\end{lstlisting}
\begin{verbatim}
# A tibble: 5 × 5
  variable effect blearner         predictor               type      
  <chr>     <dbl> <chr>            <chr>                   <chr>     
1 X18       0.364 bols(X18, int... X18                     individual
2 X5        0.250 bols(X1, X2, ... X1, X2, X3, X4, X5      group     
3 X15      -0.249 bols(X11, X12... X11, X12, X13, X14, X15 group     
4 X4        0.234 bols(X1, X2, ... X1, X2, X3, X4, X5      group     
5 X11      -0.228 bols(X11, X12... X11, X12, X13, X14, X15 group   
\end{verbatim}
\begin{lstlisting}
slice(get_coef(sgb_model = sgb_model)$aggregate,1:5)
\end{lstlisting}
\begin{verbatim}
# A tibble: 5 × 4
 variable effect learner               predictor                   
  <chr>    <dbl> <chr>                 <chr>                       
1 X18      0.364 bols(X18, inte...     X18                         
2 X15     -0.272 bols(X11, X12,...;    X11, X12, X13, X14, X15; X15
                 bols(X15, inte...
3 X5       0.250 bols(X1, X2,...       X1, X2, X3, X4, X5          
4 X4       0.234 bols(X1, X2,...       X1, X2, X3, X4, X5          
5 X13     -0.230 bols(X11, X12,...;    X11, X12, X13, X14, X15; X13  
                 bols(X13, inte...  
\end{verbatim}
We see that the effect sizes differ between the two perspectives. The variable X15, for example, has a more extreme model coefficient of -0.272 in the aggregated case compared to the coefficient of -0.249 derived only from the group base-learner. Consequently, the ordering also differs. X11 has a greater absolute model coefficient from the group than X13, but in the aggregated version, the absolute model coefficient of X13 exceeds the one of X11.
\subsubsection{Plotting model coefficients and importance}

With \verb|plot_effects()| we can plot the effect sizes of the sparse-group boosting model in relation to the relative importance to get an overall picture of the model. Through the parameter \verb|'plot_type'| one can choose the type of visualization. \verb|'radar'| refers to a radar plot using polar coordinates.
Here, the angle is relative to the cumulative relative importance of predictors, and the radius is proportional to the effect size. \verb|'clock'| does the same as \verb|'radar'|  but uses clock coordinates instead of polar coordinates. \verb|'scatter'| uses the effect size as the y-coordinate and the cumulative relative importance as the x-axis in a classical Scatter plot.
\begin{lstlisting}
plot_effects(sgb_model = sgb_model, n_predictors = 5,
             base_size = 10)
plot_effects(sgb_model = sgb_model, n_predictors = 5,
             plot_type = "clock", base_size = 10)
plot_effects(sgb_model = sgb_model, n_predictors = 5, 
             plot_type = "scatter", base_size = 10)
\end{lstlisting}
\begin{figure}[H]
\centering
\includegraphics[scale=0.58]{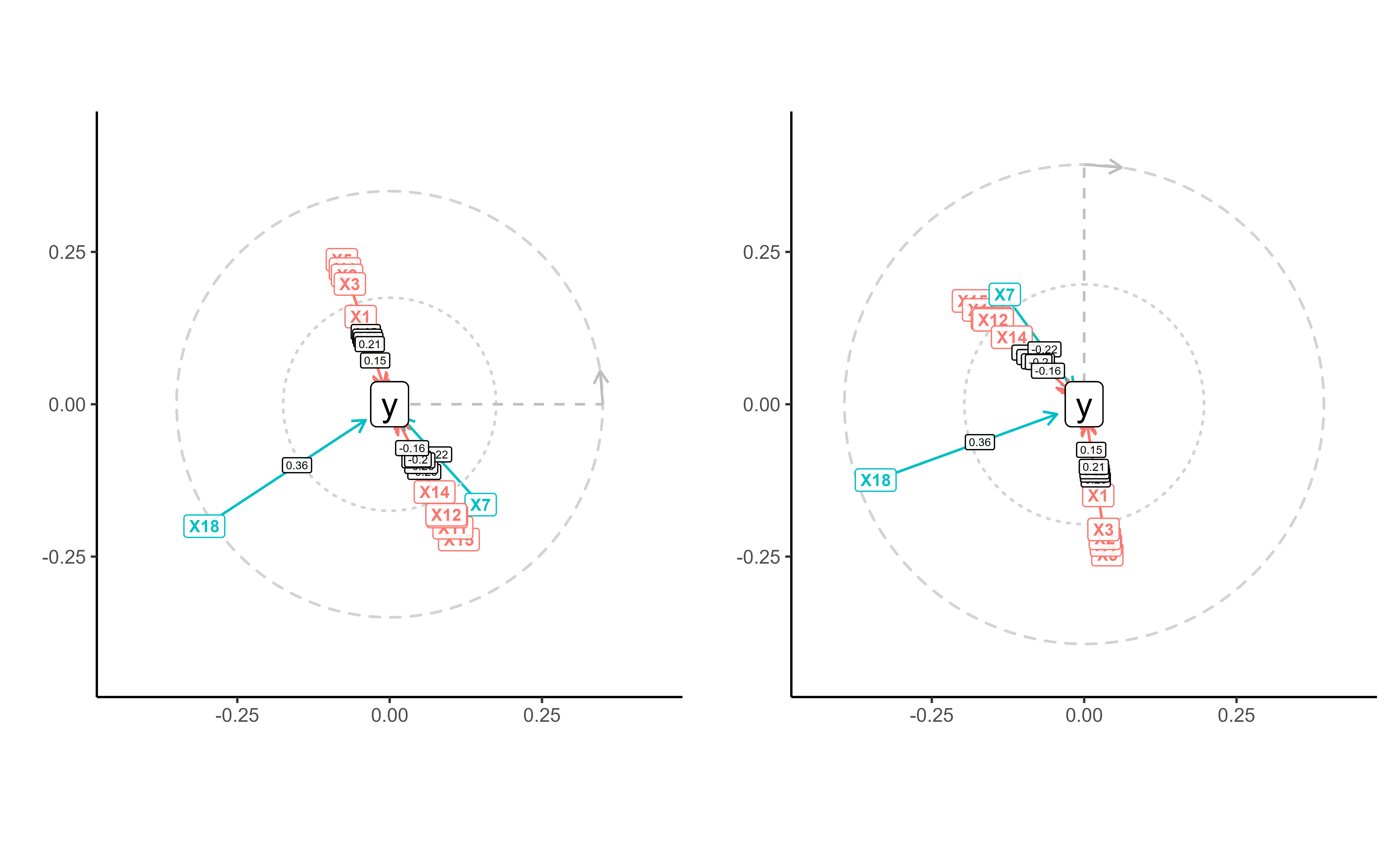} \\
\includegraphics[scale=0.58]{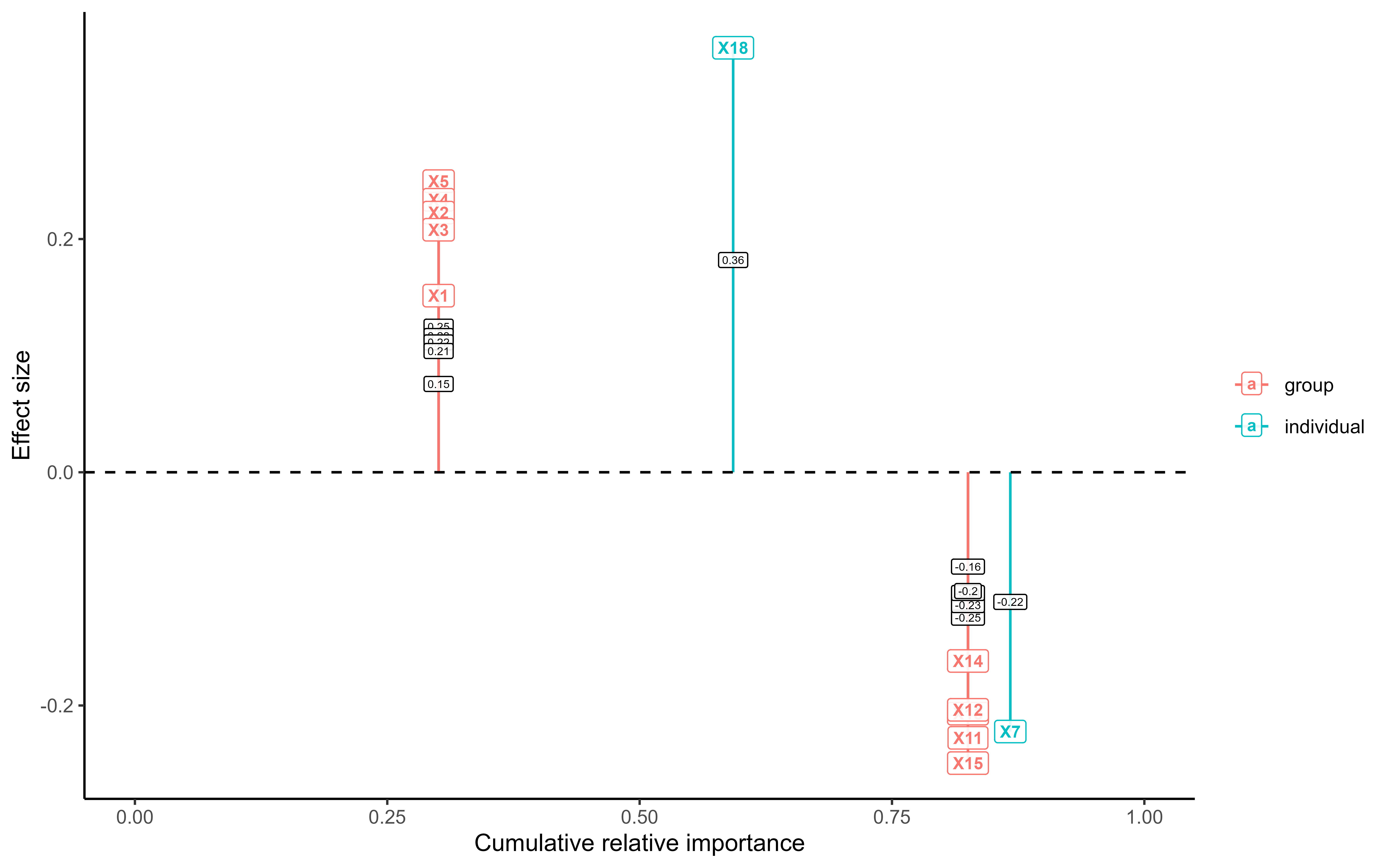}
\caption{Three visualizations of Effect size vs. relative importance of individual and group base-learners}\label{fig:radar}
\end{figure}

\subsubsection{Coefficient path}

\verb|plot_path| calls \verb|get_coef_path()| to retrieve the aggregated coefficients from a \verb|mboost| object for each boosting iteration and plots it, indicating if a coefficient was updated by an individual variable or group.
\begin{lstlisting}
plot_path(sgb_model = sgb_model)
\end{lstlisting}
\begin{figure}[H]
\centering
\includegraphics[scale=0.7]{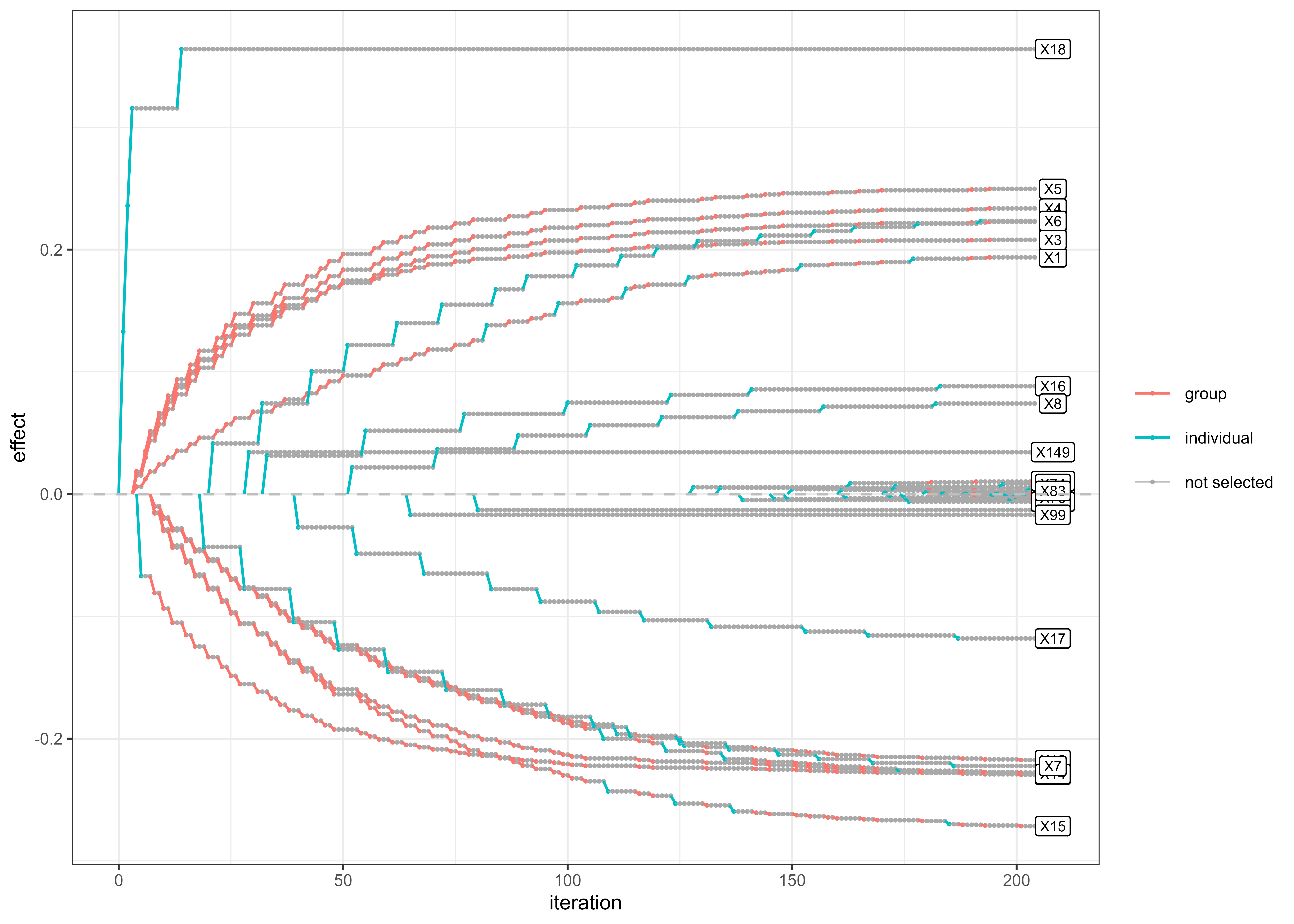}
\caption{Coefficient path of a sparse-group boosting model with simulated data}\label{fig:path}
\end{figure}
In the coefficient path shown in Figure \ref{fig:path}, we see the change in model coefficients. Since the path shows the aggregated model coefficients, the path of one variable in the dataset may have both colors. This is the case with variable X1 which was first updated through the group and then also as an individual variable or with variable X15 in reverse order.

\subsection{Real data}\label{sec3}
\begin{figure}[H]
\centering
\includegraphics[scale=0.7]{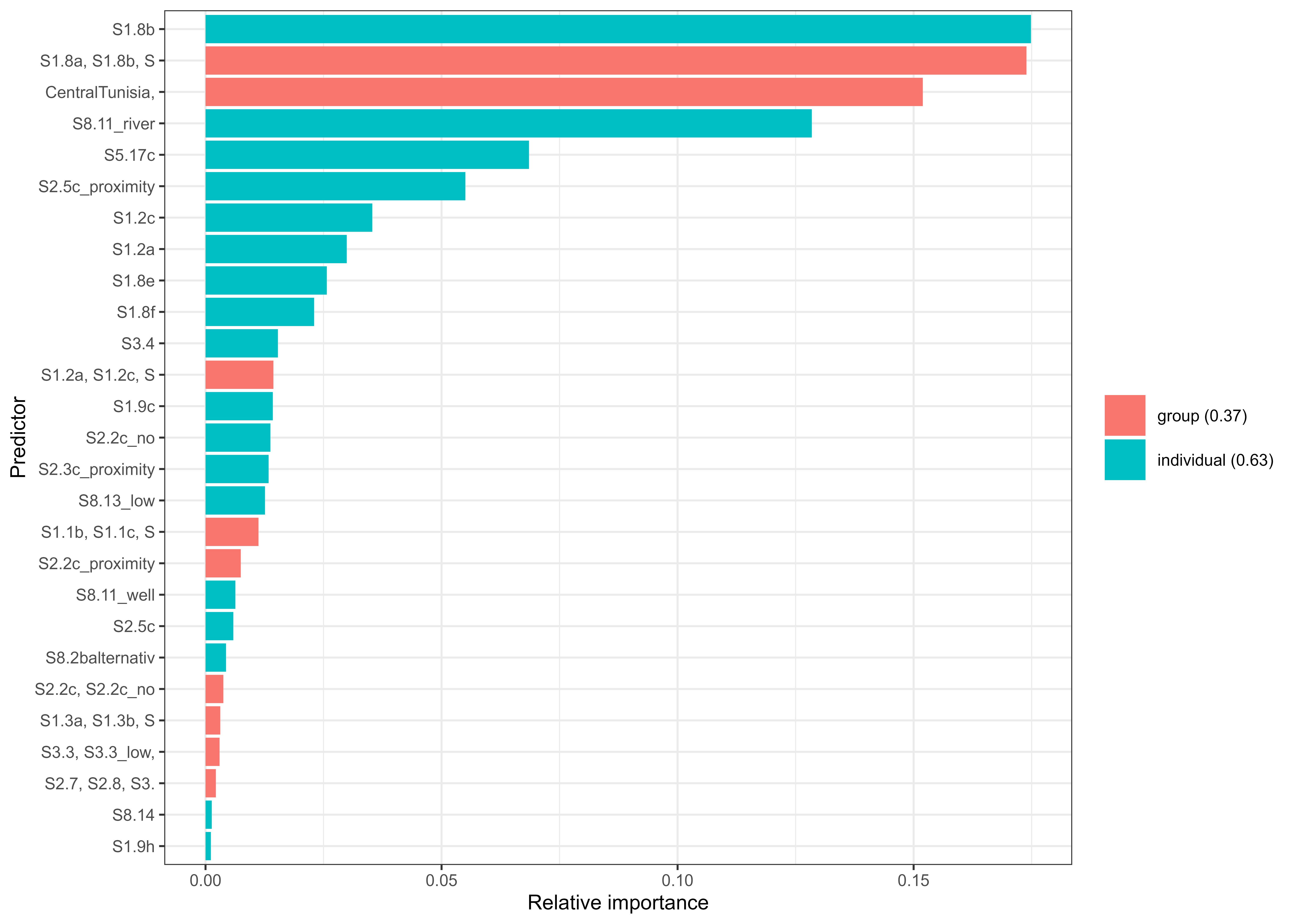}
\caption{Variable importance using the ecological dataset}\label{fig:cc_varimp}
\end{figure}

In this section, we will fit a sparse-group boosting model with \textbf{sgboost} to a real dataset. We will use behavioral ecological data and an associated group structure \cite{obster_using_2023} to explain whether farmers in Chile and Tunisia are planning to take adaptive measures against climate change in the following years. We will use a logistic regression model for this binary decision. The data consists of 14 groups and 84 variables for the 801 farmers. Groups include vulnerability to climate change \cite{pechan_reducing_2023}, social, biophysical, and economic assets, as well as perceptions of the farmers. After loading the data and group structure, we create the formula with mixing parameter $\alpha = 0.3$. Then, we pass the formula to mboost() with 1000 boosting iterations and a learning rate of 0.3. 
\begin{lstlisting}
model_df <- readRDS('model_df.RDS') %>%
  mutate_at(index_df$col_names, factor)
index_df <- readRDS('index_df.RDS')
sgb_formula <- create_formula(
  group_df = index_df, var_name = 'col_names',
  group_name = 'index', outcome_name = 'S5.4'
)
model <- mboost(
  sgb_formula, data = model_df, 
  family = Binomial(link = 'logit'),
  control = boost_control(mstop = 1000, nu = 0.3)
)
cv_model <- cvrisk(model)
model <- model[mstop(cv_model)]
\end{lstlisting}
The model is stopped early after 466 boosting iterations. We examine the coefficient path and see that in the early stage, individual base-learners were dominantly selected like the variable 'S1.8b or 'S8.11 river' which indicates whether river irrigation is used. Many of the variables were first included as individual variables and later also through group base-learners like 'S8.1b' or 'S2.5c proximity' (Proximity to extreme weather events), which we also saw in the simulated data.
\begin{lstlisting}
plot_path(model)
\end{lstlisting}
\begin{figure}[H]
\centering
\includegraphics[scale=0.7]{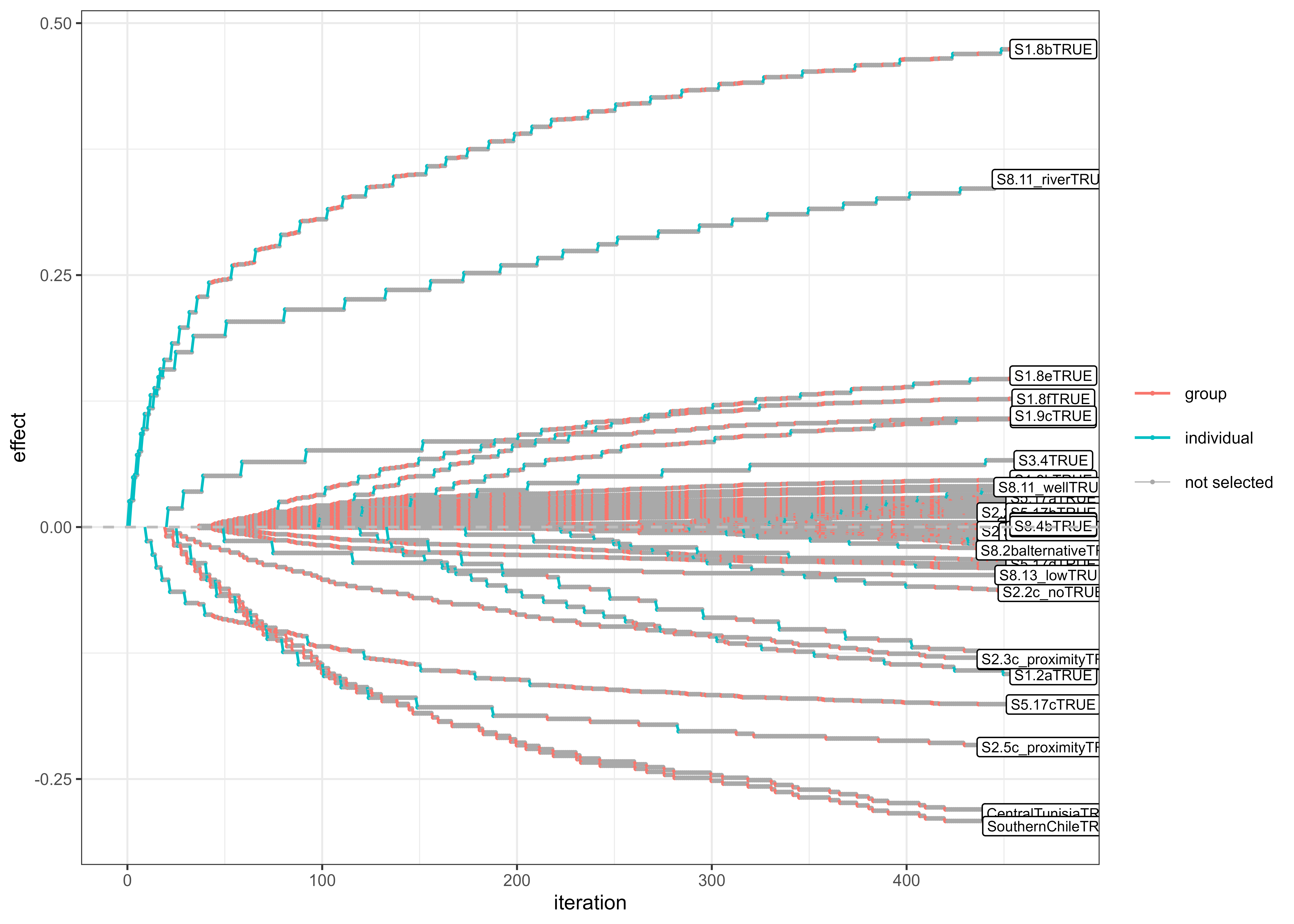}
\caption{Coefficient path using the ecological dataset}\label{fig:cc_path}
\end{figure}
In figure \ref{fig:cc_varimp}, we look at the variable importance with the default values, plotting all 27 selected predictors of which 8 are groups, the latter having a relative variable importance of 22 percent. The most important base-learner is the individual variable 'S1.8b', indicating whether farming journals are being used and the most important group is the social asset group, followed by the group consisting of the four considered regions. 
\begin{lstlisting}
plot_varimp(model)
\end{lstlisting}
Plotting the effect sizes of all predictors having a relative importance of greater than 1.5
percent shows the tendency for more important variables to have greater absolute effect sizes. For readability, we set the number of printed characters per variable to 6 and use the 'scatter' version of the plot.
\begin{lstlisting}
plot_effects(
  model, plot_type = 'scatter', 
  prop = 0.015, max_char_length = 6
)
\end{lstlisting}
\begin{figure}[H]
\centering
\includegraphics[scale=0.7]{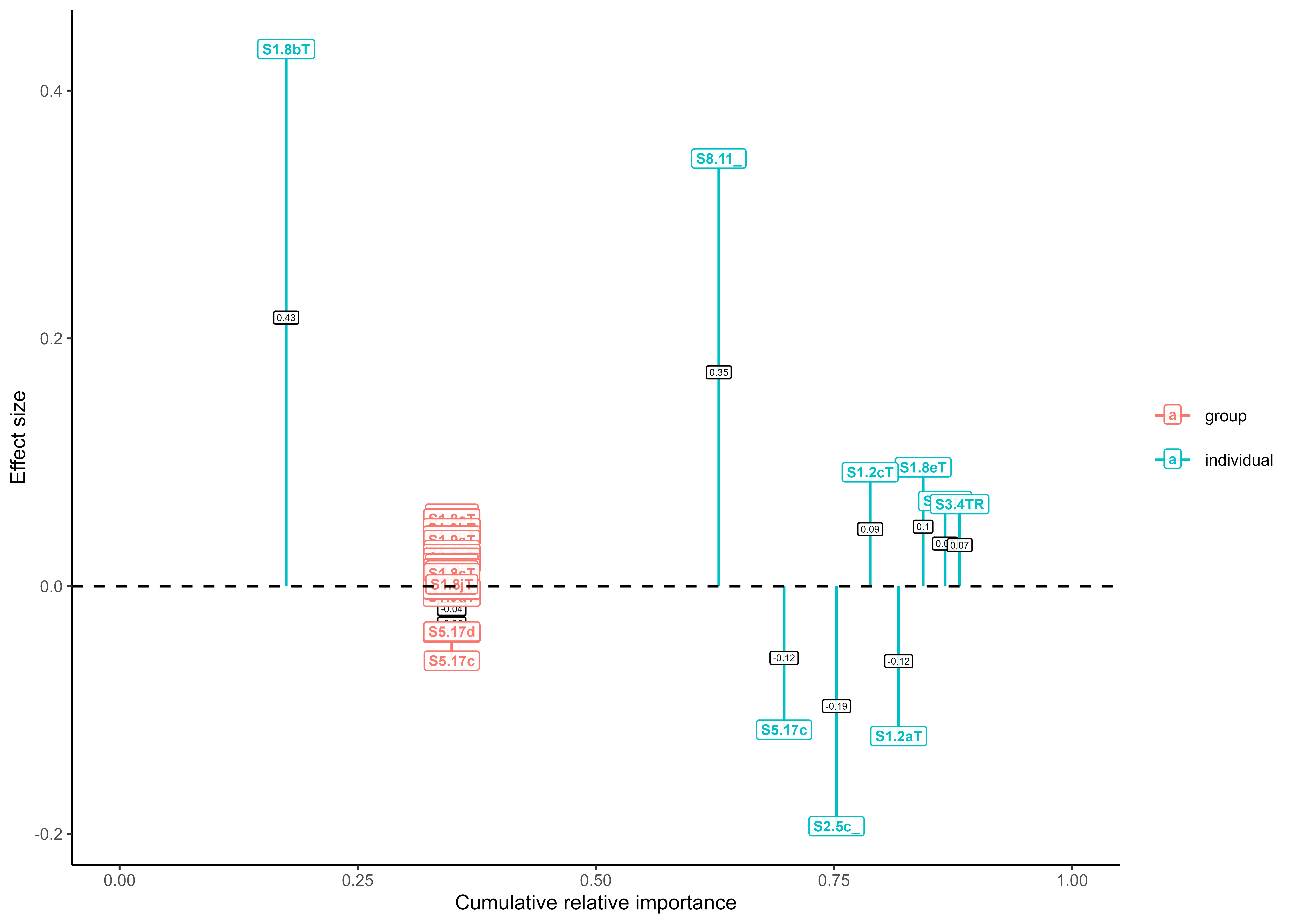}
\caption{Coefficient plot using the ecological dataset}\label{fig:cc_effects}
\end{figure}

\subsection{Group selection bias and balance}
The function \verb|balance()| returns the optimal degrees of freedom for each baselearner, such that all groups have equal selection chances if the outcome is not associated with any group. This is especially important in genetic research as group sizes based on genes may vary strongly. Not controlling for group sizes may lead to strong false detection because of a bias towards more complex groups eg. genes. To illustrate this problem, four scenarios are considered. The first three scenarios have three groups, where the first group is a categorical predictor with three categories, the second group is a categorical predictor with two categories, and the third group is a numerical variable simulated with a standard normal distribution. The fourth scenario consists of two groups, the first having 46 members and the second having 4. In the first scenario, the sample size is 50; in the second and third scenarios, the sample size is 500; in the fourth scenario, the sample size is 30, leading to $p>n$. In scenario 3 the outcome variable is i.i.d gamma distributed with shape one and rate 1, Scenarios one, two, and four have standard normal outcomes. Table 1 shows the selection frequencies in these scenarios for each group, for group boosting with three versions of group boosting: one with equal penalties ($\lambda = 0.1$), one with equal degrees of freedom df($\lambda$) = 0.5, and one with the degrees of freedom based on the \verb|balance()| function which implements Algorithm 2. The default settings of 3000 repetitions of i.i.d standard normal outcomes, 20 iterations, a learning rate of 0.5, and a reduction factor of 0.9. \\
The results suggest, that generally ridge regression with equal penalties leads to the greatest group imbalance, and equal degrees of freedom lead to relatively lower imbalances, especially when two categorical predictors are compared, and the only group adjustment-based group boosting balances the the selection frequencies in all scenarios. In scenario 4, equal lambda only selects the larger group, and in equal degrees of freedom, the chance of the larger group being selected is 2.9 times higher than the smaller one. equal lambda is better at balancing a binary variable with a numerical variable compared to equal df, where equal df is better at balancing group sizes compared to equal lambda. The distribution of the outcome variable seems to not play a great role as the results in scenarios two and three are quite comparable for all three models, which makes the balancing algorithm robust, even if the distribution of the error is not known. Comparing scenario one with two, the balancing algorithm works better with a greater sample size. 
\begin{table}[ht!]
\caption{Selection frequencies in group boosting with degrees of freedom adjustment (group adjustment) compared to ridge regression with equal degrees of freedom (equal df) and equal penalty term (equal lambda). The degrees of freedom used as group adjustment are shown in brackets.}
  \begin{center}
    \label{table:sim_params}
     \begin{tabular}{c | c c c c }
      Scenario & group & equal lambda & equal df & group adjustment \\
       \hline \hline
        1 & 1 & 0.699 & 0.453 & 0.345 (df=0.377) \\
        \hline
        1 & 2 & 0.157 & 0.364 & 0.341 (df=0.406) \\
        \hline
         1 & 3 & 0.144 & 0.183 & 0.314 (df=0.717) \\
        \hline
         2 & 1 & 0.701 & 0.407 & 0.337 (df=0.397) \\
        \hline
        2 & 2 & 0.15 & 0.419 & 0.339 (df=0.352) \\
        \hline
         2 & 3 & 0.149 & 0.174 & 0.324 (df=0.751) \\
        \hline
        3 & 1 & 0.695 & 0.417 & 0.338 (df=0.397) \\
        \hline
        3 & 2 & 0.155 & 0.408 & 0.326 (df=0.352) \\
        \hline
        3 & 3 & 0.15  & 0.175 & 0.336 (df=0.751) \\
        \hline
        4 & 1 & 1 & 0.744 & 0.518 (df=0.394)\\
        \hline
        4 & 2 & 0 & 0.256 & 0.482 (df=0.606)\\
\end{tabular}
  \end{center}
\end{table}
\FloatBarrier
\section{Discussion}
\subsection{Sparse-group boosting}
'sgboost' when applied to high-dimensional grouped data such as ecological data on climate adaptation, reveals meaningful patterns, such as socio-economic and biophysical variables, thereby providing actionable insights for policy and practice. Moreover, the integration of comprehensive visualization tools—such as variable importance plots, coefficient paths, and effect size charts—enhances the interpretability of the models, making it easier for practitioners to understand the contributions of different predictors.
\subsection{Group bias}
Attempts to reduce selection bias in boosting have been made through equal degrees of freedom using the definition $\text{df}(\lambda) = \text{tr}(2H^\lambda-(H^\lambda)^2)$ \cite{hofner_framework_2011}
The results of our simulations highlight a fundamental issue in group boosting: a systematic selection bias towards more complex base-learners. This phenomenon follows the principle that "whoever shouts the loudest is rarely right," meaning that larger or more flexible groups are favored in the selection process.  The proposed group balancing function \verb|balance()| in sgboost implementing the group balancing algorithm effectively eliminating this bias by using a simulation-based approach to equalize selection probabilities across different groups. \\
While this issue is particularly evident in group boosting, it is not limited to this setting. A similar bias can arise in standard boosting when base-learners differ significantly in scale or distribution. This occurs, for example, when comparing binary and numerical variables or in the context of functional regression. The proposed adjustment method provides a systematic way to address these imbalances across various modeling settings and is robust against small sample sizes and varying outcome variable distributions. Even if one assumes a wrong error distribution in the simulation e.g. standard normal, the group adjustment seems to still balance the group selection frequencies if the actual error distribution differs, e.g. gamma errors. 

\subsection{Limitations of the group balancing algorithm}
Despite its benefits, the balancing approach has several challenges: The resampling and iterative adjustments make the method time-intensive. Different combinations of degrees of freedom can yield similar selection frequencies. One way to address this is by fixing the degrees of freedom of a reference base-learner and adjusting only the others.
If the learning rate is too high or the number of resamples is too low, the algorithm may fail to converge. To mitigate this, a reduction factor is applied.
The method can overshoot adjustments, leading to situations where the ridge regression problem becomes ill-posed (e.g., degrees of freedom approaching zero or leading to negative. This is controlled via predefined bounds (\verb|max_df| and \verb|min_df|) which can be used instead of the too-extreme solution. \\
While the algorithm increases computational cost, it is important to compare this to alternative approaches. Standard group boosting with equal penalties often requires extensive tuning, including thousands of boosting iterations and 25-fold cross-validation to determine optimal stopping. In contrast, the balancing approach achieves the same computational efficiency when considering this tuning overhead. Furthermore, parallelization can significantly reduce runtime. Additionally, equal degrees of freedom approaches also require multiple 
$\lambda$ values to equalize the selection behavior, making the balancing method computationally more efficient in comparison if the $\lambda$ values are optimized and not the degrees of freedom.
\section*{Funding}

This research is funded by dtec.bw – Digitalization and Technology Research Center of the Bundeswehr. dtec.bw is funded by the European Union – NextGenerationEU. All statements expressed in this article are the authors’ and do not reflect the official opinions or policies of the authors’ host affiliations or any of the supporting institutions.

\bibliographystyle{tfs}
\bibliography{refs}

\end{document}